\documentclass[american,aps,prx,superscriptaddress,reprint]{revtex4-1}
\usepackage{graphicx,color}  
\usepackage{wrapfig}
\usepackage{dcolumn}   
\usepackage{bm}        
\usepackage{amssymb}   
\usepackage{amsmath} 
\usepackage{subfig}
\usepackage[unicode=true,
 bookmarks=true,bookmarksnumbered=false,bookmarksopen=false,
 breaklinks=false,pdfborder={0 0 0},pdfborderstyle={},backref=false,colorlinks=true]
 {hyperref}
\hypersetup{
 pdfborderstyle={},allcolors=blue}

\begin{document}


\title{Enhancing quantum control by improving shape pulse generation}

\author{John P. S. Peterson}

\thanks{e-mail:  johnpetersonps@hotmail.com}

\affiliation{Institute for Quantum Computing and Department of Physics and Astronomy,
University of Waterloo, Waterloo N2L 3G1, Ontario, Canada}
                        
\author{Roberto S. Sarthour}

\affiliation{Centro Brasileiro de Pesquisas Físicas, Rua Dr. Xavier Sigaud 150,
22290-180 Rio de Janeiro, Rio de Janeiro, Brazil}

\author{Raymond Laflamme}

\affiliation{Institute for Quantum Computing and Department of Physics and Astronomy,
University of Waterloo, Waterloo N2L 3G1, Ontario, Canada}

\affiliation{Perimeter Institute for Theoretical Physics, 31 Caroline Street North, Waterloo, Ontario, N2L 2Y5, Canada}

\affiliation{Canadian Institute for Advanced Research, Toronto, Ontario M5G 1Z8, Canada}
\date{\today}

\begin{abstract}
Most quantum processors requires pulse sequences for controlling quantum states. Here, we present an alternative algorithm for computing an optimal pulse sequence in order to perform a specific task, being an implementation of a quantum gate or a quantum state preparation. In our method, we reduced drastically the number of parameters to be fitted, by using a limited number of functions as the modulations for the amplitude and phase of the radio-frequency pulses, and employed approximations to make the algorithm fast and scalable. We demonstrate the success of the proposed algorithm, by performing several real experiments for 4, 7 and 12 quantum bits systems using NMR. In addition, we have also shown the efficiency of the algorithm, finding pulses for controlling with good fidelity the quantum states of spins in a fictional square bi-dimensional lattices containing 16, 36 and 100 qubits.  
\end{abstract}

\pacs{}
\maketitle


\section{Introduction}

Precise control of quantum systems is necessary for many important experimental implementations in the areas of quantum information and quantum computation. Furthermore, the control must be faster when compared to decoherence times, which is paramount for quantum computing \cite{revisao,livro,livronc}. In general, the theoretical plans demand that external operations – mainly electric and magnetic time dependent fields – or internal interactions to be turned on and off as quick as possible. These processes are the ones used to execute the quantum gates \cite{revisao}. Although many of these operations can be achieved very rapidly, in practice, small deviations are likely to affect profoundly the experimental results, since many of these processes are required for simulating or executing a quantum protocol. In experiments where high accuracy is necessary, these implementation errors can significantly affect the final results.   The method to obtain the optimal conditions for performing the quantum gates is to use numerical calculations that can account for the imperfections and delays in the experimental apparatuses. Furthermore, it is possible to use numerical simulations to find the optimal forms of the external excitations needed for controlling the quantum systems. Currently, in a modern NMR (Nuclear Magnetic Resonance) equipment, it is possible to implement magnetic radio-frequency pulses with amplitude and phase modulations, which are used to construct the quantum gates and/or to perform quantum simulations. These radio-frequency pulses are, in general, developed to be robust to various types of errors, such as calibration errors, relaxation and variations of the resonance frequency. One of the main algorithms used to obtain these modulated pulses is GRAPE, Gradient Ascent Pulse Engineering \cite{grape}. Not only in NMR, GRAPE algorithms are being applied in many others experimental techniques and even in different areas \cite{epr,nvcenter,iontrap,circuit,revisao}. In the NMR case, the optimization method consists of dividing the radio-frequency (RF) excitation into many small intervals of time, short pulses, and for each one fit parameters (amplitude and phase) that will give the best description of the desired unitary operation. In fact, others numerical methods of optimization were also implemented \cite{funcao1,funcao2,krotov,Lyapu,chop1,chop2,goat,grapeexp,compila}, attempting to ease the computational processes that are hard due to the high number of parameters which must be taken into account. The number of parameters to be fitted is high because it is typically necessary to use tenths of pulses, and sometimes hundreds. Constraints are also necessary, since it is hard for the NMR spectrometer to cope with rapid variations between the RF pulses, and thus smooth changes, from pulse to pulse, are required. This limits the search for the best parameters. These numerical calculations are in general very hard and take lots of computational time to be executed. In some cases, even after many computing hours a satisfactory solution is not achieved. This is due mainly to the size of the system, number of parameters and equipment limitations. Although, some progress has been made there still a long way to go.

In this work, we present an algorithm to optimize external excitations that are used to manipulate the quantum states of relatively large systems. For this, only a reasonable small number of parameters to be optimized are necessary. In addition, our algorithm was designed to work with some approximations to make it fast and scalable. Therefore, the time to perform the optimization is drastically reduced. We demonstrate the success of the proposed method by applying our algorithm in several real experiments, in which we manipulate the quantum states of NMR systems containing 4, 7 and 12 qubits. Furthermore, we have also shown the efficiency of the algorithm, finding pulses for controlling with good precision the quantum state of spins distributed in a fictional bi-dimensional square lattices containing 16, 36, 100 qubits. At the end, we have also discussed the application of our method in a larger system containing 65536 qubits. 

\section{Optimal control theory}

When we use a quantum system to perform a quantum computation or simulate the dynamics of other physical systems, generally, we have to implement a unitary operator ($U_{goal}$) that is not possible to be produced only with the natural evolution of the system at hand. Hence, we need to add external interactions in order to modify the natural dynamics of the system. If we include these interactions, the total Hamiltonian of the system will be given by:
\begin{equation}\label{eq:ht}
 \begin{split}
\ \mathcal{H}_{T}(t) = \mathcal{H}_{0} + \mathcal{H}_{C}(t),
 \end{split}
 \end{equation} 
where $\mathcal{H}_{0}$ represents the natural Hamiltonian of the system and $\mathcal{H}_{C}(t)$ is the control Hamiltonian that describes the interactions used to modify the natural dynamics of the system.
 
The evolution of the system under the action of the Hamiltonian $\mathcal{H}_{T}(t)$ will produce the following unitary:
\begin{equation}\label{eq:uht}
 \begin{split}
\ U_{\mathcal{H}_{T}} = \mathcal{T} \left[ \exp\left(-\frac{i}{\hbar}  \int \mathcal{H}_{T}(t) dt \right) \right].
 \end{split}
 \end{equation} 
In Eq.~\eqref{eq:uht}, $\mathcal{T}$ represents the Dyson time-ordering operator and $\hbar$ is the Planck constant divided by $2\pi$. In order to have $U_{\mathcal{H}_{T}} = U_{goal}$, we must find the values of $\mathcal{H}_{T}(t)$ that minimize the function
\begin{equation}\label{eq:fide}
 \begin{split}
\ \mathcal{F} = 1 - \frac{\left | Tr\left( U_{goal}^{\dagger }U_{\mathcal{H}_{T}}\right)  \right |}{N},
 \end{split}
 \end{equation} 
where $N$ is the dimension of the Hilbert space. The global minimum of $\mathcal{F}$ can be very hard to find, since the number of operations needed in the control Hamiltonian, which are necessary to get $U_{\mathcal{H}_{T}} = U_{goal}$, are usually very high. However, using numerical optimizations, we can find local minima where $U_{\mathcal{H}_{T}}\approx U_{goal}$.

To perform a numerical optimization, the time must be discretized in $m$ intervals of duration $\delta t$. The value of $\delta t$ must be small enough to allow us to consider that $\mathcal{H}_{T}(\delta t)$ is reasonably constant at each of the $m$ intervals. In this case, we can calculate $U_{\mathcal{H}_{T}}$ using the following equation: 
\begin{equation}\label{eq:uhtpro}
 \begin{split}
\ U_{\mathcal{H}_{T}} = U_{m}U_{m-1}U_{m-2} \cdots U_{2}U_{1}, 
 \end{split}
 \end{equation} 
with
\begin{equation}\label{eq:uhtprok}
 \begin{split}
\ U_{k} =  \exp\left\lbrace-\frac{i}{\hbar} \left[ \mathcal{H}_{0} + \mathcal{H}_{C}\left( k\delta t \right) \right]  \delta t \right\rbrace .
 \end{split}
 \end{equation}
Nowadays, we know many algorithms that can be used to find local minima of $\mathcal{F}$ \cite{livrooti}. The choice of one of these algorithms is usually based on the form of $\mathcal{H}_{T}(t)$. In our case, $\mathcal{H}_{T}(t)$ will be the Hamiltonian of a system of nuclear spins that are controlled using the NMR technique.

\section{NMR}

In the NMR technique, a sample containing many molecules, whose elements  have nuclear spins, is placed in a uniform magnetic field along the $z$-direction, and radio-frequency pulses are used to control the quantum states of these spins. For a homonuclear NMR system, all the spins of interest have the same gyromagnetic ratio \cite{livrole}, since they are all of the same kind, and they are subject to the same magnetic field along the $z$-direction. However, due to the electrons clouds of the neighbour atoms, each individual nuclear spin is subject to a slightly different magnetic field. This is known as chemical shift \cite{livrole}. In addition, there is also the coupling interaction between the spins, which occurs via the exchange mechanism. Generally, the samples we utilize in order to simulate quantum systems or implement algorithms are isotropic liquids. The control of such systems is easier, since several interactions do not significantly influence the dynamics of these systems \cite{livrole}. Thus, we will consider samples that can be described by the following Hamiltonian:
\begin{equation}\label{eq:h0}
 \begin{split}
\ \mathcal{H}_{0} = \sum_{k}\frac{\hbar(\omega_{k}-\omega_{R})\sigma_{z_{k}}}{2} + \sum_{k \neq n}\frac{\pi\hbar J_{kn}\sigma_{z_{k}}\sigma_{z_{n}}}{4},
 \end{split}
 \end{equation} 
where $\omega_{k}$ and $\sigma_{\beta_{k}}$ are, respectively, the angular resonance frequency and the Pauli matrix $\beta$ of the $k$-th nuclear spin, $\omega_{R}$ is the angular frequency of the rotating frame and $J_{kn}$ is the scalar coupling constant of the spins $k$ and $n$. 
 
For controlling the quantum states of the nuclear spins, we utilize radio-frequency pulses applied in the $xy$ plane with an angular frequency $\omega_{R}$. The interactions of the spins with a pulse can be described by the following Hamiltonian:
\begin{equation}\label{eq:hc}
 \begin{split}
\ \mathcal{H}_{C}(t) = \hbar \Omega(t) \sum_{k}\frac{\cos[ \phi(t) ] \sigma_{x_{k}} + \sin[ \phi(t) ] \sigma_{y_{k}}}{2},
 \end{split}
 \end{equation} 
where $\Omega(t)$ and $\phi(t)$ represent the modulations of the
pulse amplitude and its phase.

\section{The Algorithm}

According to Fermi’s Golden Rule, time dependent operations are necessary for inducing transitions between different energy levels. Experimentally, these
transitions can be induced through the use of an oscillating electromagnetic field. In this way, electromagnetic fields can be used to implement quantum gates. The precise tailoring of these quantum gates is done through the modulation of the phase and amplitude of these fields. When the phase and amplitude do not have high symmetry, the optimal values for implementing a particular quantum gate must be found numerically. The current standard technique for finding these parameters is the GRAPE algorithm \cite{grape}. This algorithm optimize the amplitude and phase at each time step requiring hundreds of parameters for the construction of a quantum gate, which is a computational arduous task. Furthermore, the GRAPE algorithm commonly requires many hours of computation and can, at times, result in gates with poor fidelity \cite{grapeexp}. The algorithm we propose in this work avoids processing large numbers of parameters by using a set of functions to find the shapes of the amplitude and the phase for the pulse. As a result, the number of parameters that need to be determined numerically are drastically reduced. For illustrative purposes, we present a formulation of our algorithm for finding pulses for the NMR technique. However, this algorithm can be extended to other techniques which use electromagnetic pulses for the control of a quantum system.

Since a Fourier series can be used to describe any function, we have chosen a limited series of sinusoidal functions where the amplitudes, frequencies and phases have to be fitted. The functions produced by this fit will be the envelop of the amplitude and phase of the radio-frequency pulse. Thus, the amplitude and the phase of the pulse are modulated using sums composed of $s_{A}$ and $s_{P}$ sines, respectively.
\begin{equation}\label{eq:omega}
 \begin{split}
\ \Omega(t) = \sum_{k=1}^{s_{A}} a_{k}\sin \left( b_{k}t + c_{k} \right),
 \end{split}
 \end{equation} 
\begin{equation}\label{eq:phi}
 \begin{split}
\ \phi(t) = \sum_{k=1}^{s_{P}} d_{k}\sin \left( f_{k}t + g_{k} \right).
 \end{split}
 \end{equation} 
The variables $a_{k}$, $b_{k}$, $c_{k}$, $d_{k}$, $f_{k}$ and $g_{k}$ must be optimized in order to obtain $U_{\mathcal{H}_{T}}\approx U_{goal}$. Generally, at the end of the optimization, the values of these variables will be of the same order of their initial value, reminding that an initial guess has to be given as an input. The values that the function $\Omega(t)$ can assume must belong to the range $[0,A_{max}]$, where the upper bound will be established by the experimental equipment. To ensure that the function $\Omega(t)$ does not exceed the lower bound, we have to shift this function so that its minimum is always positive. This can be accomplished with the following substitution: $\Omega(t) \rightarrow  \Omega(t) - min\left[ \Omega(t) \right] $. Meanwhile, to limit the maximum value of $\Omega(t)$, we must find the maximum of this function, $\Omega_{max} = max [\Omega(t)]$, and divide it by the limit of the amplitude, $\Omega_{max}/A_{max}$. If the result of this division is greater than $1$, we need to make the following substitution $\Omega(t) \rightarrow  \Omega(t)A_{max}/\Omega_{max} $.

We employ a Nelder-Mead simplex algorithm to solve the optimization problem \cite{livrooti}. This algorithm does not use derivatives of the function $\mathcal{F}$. In our case, we obtained good results using the fminsearch function from the MATLAB software. Considering the case where 63 parameters of Eq.~\eqref{eq:omega} and Eq.~\eqref{eq:phi} ($s_{A}=7$ and $s_{P}=14$) are optimized, this algorithm converged faster than the optimization method used in GRAPE, which requires derivatives of $\mathcal{F}$. 

When compared to GRAPE, another advantage we have in our method is that by increasing the pulse duration or reducing the interval $\delta t$, the number of variables that must be optimized does not increase. Consequently, the time to perform the optimization increases linearly. In our algorithm, we use this advantage to calculate the value of $U_{k}$ quickly. For this, we utilize the approximation presented in \cite{apro}, which requires a small $\delta  t$ to calculate the value of $U_{k}$ with a good precision. Thus, for a system composed of $q$ qubits, we have 
\begin{equation}\label{eq:uhtprokaproximado}
 \begin{split}
\ U_{k} \approx e^{-i\phi (k\delta t)\Gamma } W_{1} e^{-i\Omega (k\delta t)\Gamma \delta t} W_{2} e^{i\phi (k\delta t)\Gamma } ,
 \end{split}
 \end{equation}
with $\Gamma = \sum_{l=1}^{q}\sigma_{z_{l}}/2$, $W_{1}=e^{-i \mathcal{H}_{0} \delta t/2} H_{q}$ and $W_{2} =H_{q} e^{-i \mathcal{H}_{0} \delta t/2}$. The matrix $H_{q}$ is the tensor product of $q$ Hadamard gates. Note that the values of $W_{1}$ and $W_{2}$ only need to be calculated once. Since the other matrices of Eq.~\eqref{eq:uhtprokaproximado} are diagonal in the computational basis, in order to find the value of $U_{\mathcal{H}_{T}}$, we need to calculate only exponentials of numbers and matrix products. Although gradient-free optimization methods require more function evaluations, the union of our method with the approximation presented by Bhole and Jones \cite{apro} makes the average execution time of our algorithm less than the ones  obtained using GRAPE (with this approximation) for the systems presented in the results section.

\subsection{Considerations for the Pulse Amplitude}

Another common problem we have to consider are the initial and final values of the pulse amplitude. The RF generator and amplifiers have limited time response that
constraint the initial and final values of the pulse amplitude. This constraint can be solved if a smooth function $\Omega(t)$ is used, and it meets the following conditions:
 \begin{itemize}
   \item The initial and final value of the amplitude must be null;
   \item The rate of change of the amplitude, both at the beginning and at the end, have to be able to match the restriction due to equipment used.  
  \end{itemize}

In our algorithm, we were able to satisfy these conditions multiplying the amplitude of a pulse with duration $\tau_{f}$ by the function
\begin{equation}\label{eq:tanh}
 \begin{split}
\ \Lambda \left( t \right) = -\tanh \left[ \frac{ \zeta_{1} t}{ \tau_{f} } \right] \tanh \left[ \frac{ \zeta_{2} \left( t- \tau_{f} \right) }{ \tau_{f} } \right].
 \end{split}
 \end{equation}
The rate of change of the amplitude may be reduced when the values of $\zeta_{1}$ and $\zeta_{2}$ are reduced. The values of these constants are experimentally determined. It is worth mentioning that, in general, when we reduce the values of $\zeta_{1}$ and $\zeta_{2}$ the optimization becomes more challenging. Due to this fact, we should look for the highest values of $\zeta_{1}$ and $\zeta_{2}$ that produce pulses that are well implemented. With our equipment we use $\zeta_{1} = \zeta_{2} = 2$ and obtain good experimental results.

\subsection{Radio-Frequency Considerations}

In our algorithm, we can also include a condition to obtain pulses that are robust to amplitude calibration. If we include this condition, we have to optimize two more unitary operations, which are given by
\begin{equation}\label{eq:upm}
 \begin{split}
\ U_{-} = \mathcal{T} \left[ \exp\left(-\frac{i}{\hbar}  \int \mathcal{H}_{0} + (1-\varepsilon )\mathcal{H}_{C}(t) dt \right) \right],\\
\\ U_{+} = \mathcal{T} \left[ \exp\left(-\frac{i}{\hbar}  \int \mathcal{H}_{0} + (1+\varepsilon )\mathcal{H}_{C}(t) dt \right) \right]. 
 \end{split}
 \end{equation}  
The constant $\varepsilon$ represents the value of the error in the calibration of the pulse amplitude. In our tests we used $\varepsilon = 0.05$, which is equivalent to an error of $5 \%$. The new function that we have to optimize will be given by the following weighted average:
\begin{equation}\label{eq:fiderf}
 \begin{split}
\ \mathcal{F}_{RF} = \frac{ \alpha_{1} \mathcal{F}_{-} + \alpha_{2} \mathcal{F} + \alpha_{3}\mathcal{F}_{+} }{ \alpha_{1} + \alpha_{2} + \alpha_{3} },
 \end{split}
 \end{equation}
where $\alpha_{1}$, $\alpha_{2}$ and $\alpha_{3}$ are the weights of each element of this average. The values of these weights are defined experimentally. Meanwhile, the values of $\mathcal{F}_{-}$ and $\mathcal{F}_{+}$ can be calculated, respectively, replacing $U_{\mathcal{H}_{T}}$ by $U_{-}$ and $U_{+}$ in Eq.~\eqref{eq:fide}. 

In principle, caused by the calculation of the new operators $U_{-}$ and $U_{+}$, the number of operations we need to perform for obtaining the value of $\mathcal{F}_{RF}$ is three times the number to obtain $\mathcal{F}$. However, using Eq.~\eqref{eq:uhtprokaproximado}, the value of $\mathcal{F}_{RF}$ can be calculated efficiently, and the number of operations increases by approximately $50\%$. In order to achieve this improvement, during the calculation of $U_{-}$, $U_{\mathcal{H}_{T}}$ and $U_{+}$ we have to take into account that in Eq.~\eqref{eq:uhtprokaproximado} only the term that depends of $\Omega(k\delta t)$ will change.

In modern NMR equipment, the errors in the amplitude calibration of the pulse are higher than the errors in the phase calibration \cite{manual}. Thus, we do not include conditions for the pulses to be robust to phase calibration errors.

\subsection{Resonance Frequency}

As it was done for the error in the pulse amplitude, we can consider errors in the resonance frequency. In this case, to obtain the operators $U_{-}$ and $U_{+}$ we must multiply, respectively, $\omega_{k}$ by $(1-\varepsilon )$ and $(1+\varepsilon )$, instead of multiplying $\mathcal{H}_{C}(t)$. Therefore, if we use Eq.~\eqref{eq:uhtprokaproximado}, only the matrices $W_{1}$ and $W_{2}$ will be modified in the calculation of $U_{-}$ and $U_{+}$. In our experiments we did not include this condition, because the pulses obtained were already robust to this type of error. As an example, in a system with 4 qubits we observed that even altering the frequency of resonance up to $30$ Hz, the value of $\mathcal{F}$ remains less than 0.001.

\begin{figure}[b]%
    \includegraphics[width=8.5cm]{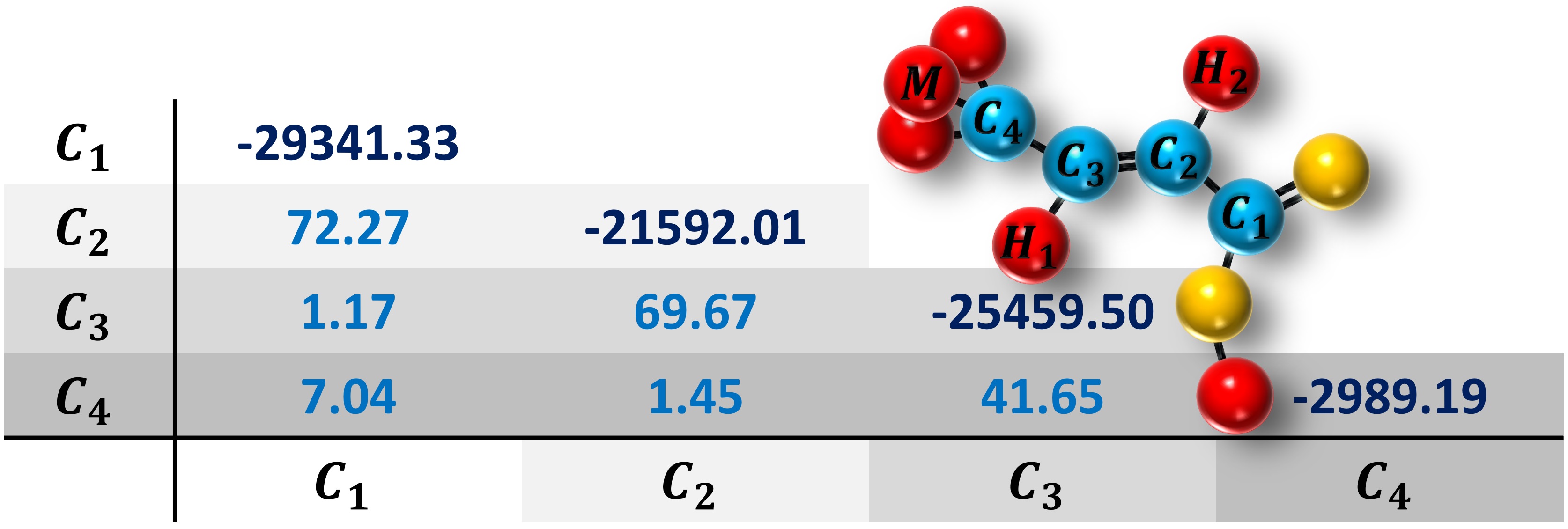}
    \centering
    \caption{ Sample information for $^{13}\textrm{C}$-labelled transcrotonic acid molecule - The off-diagonal terms in the table are the $J$ coupling constants of the $^{13}\textrm{C}$ nuclear spins of the $^{13}\textrm{C}$-labelled transcrotonic acid molecule. Meanwhile, on the diagonal we have the values of the chemical shifts of each nuclear spin. The values in the table are in Hz.}%
    \label{fig:molecula4q}%
\end{figure}

\section{Experimental Results}

For demonstration purposes, we have performed some calculations using the algorithm and used these results in real NMR experiments. Here, we present examples where the amplitude and the phase of the pulses were optimized in order to implement some quantum gates. The pulses are optimized from a random initial guess. Although this algorithm can be used to optimize gates of two qubits or more, we recommend the programmer to decompose such gates into free evolutions under the Hamiltonian $\mathcal{H}_{0}$ and gates of one qubit. By decomposing these gates, we can reduce the amount of errors, since pulse calibration errors do not occur during free evolutions. Furthermore, the time for optimizing the pulse sequence which implements the quantum gates will also be shorter. Following this approach, the errors due to the free evolutions can be diminished using refocus sequences \cite{refoco}, or the method presented by Ryan et al \cite{compila}, or an optimization method like the one presented in \cite{oti}. In our tests, we optimized the pulses to implement sequences used to prepare the pseudo-pure state (PPS) \cite{livro} for a system with 4 and 7 qubits, and some pulses to control a system of 12 qubits. After the numerical optimization, we performed the experiments using a Bruker Avance III $700$ MHz NMR spectrometer. All the experiments were performed at room temperature.

\begin{figure*}[t!]%
    \includegraphics[width=17.7cm]{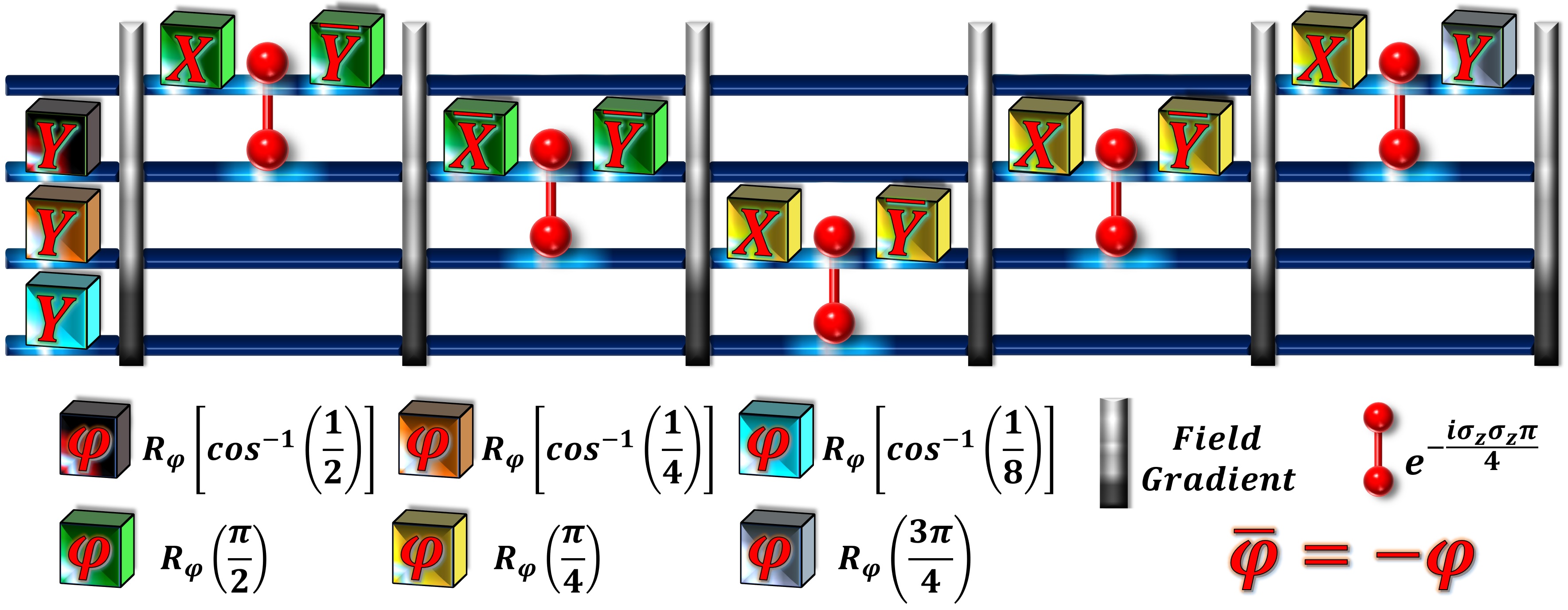}
    \centering
    \caption{ Quantum circuit to prepare the pseudo-pure state $\left | 1111  \right \rangle$ from a thermal state - In addition to rotations, we use magnetic field gradients along the $\widehat{z}$ direction and free evolution. The gate $\exp \left ( -i\sigma_{z}\otimes \sigma _{z} \pi /4 \right )$ can be implemented using two free evolutions and $\pi$ rotations, in the target qubits of this gate, after each evolution. The time of these evolutions is equal to $1/4J_{kn}$, where $J_{kn}$ is the scalar coupling constant of the target qubits.}%
    \label{fig:pps4}%
\end{figure*}

\begin{figure*}[t!]%
    \centering
    \subfloat[Amplitude]{{\includegraphics[width=8.5cm]{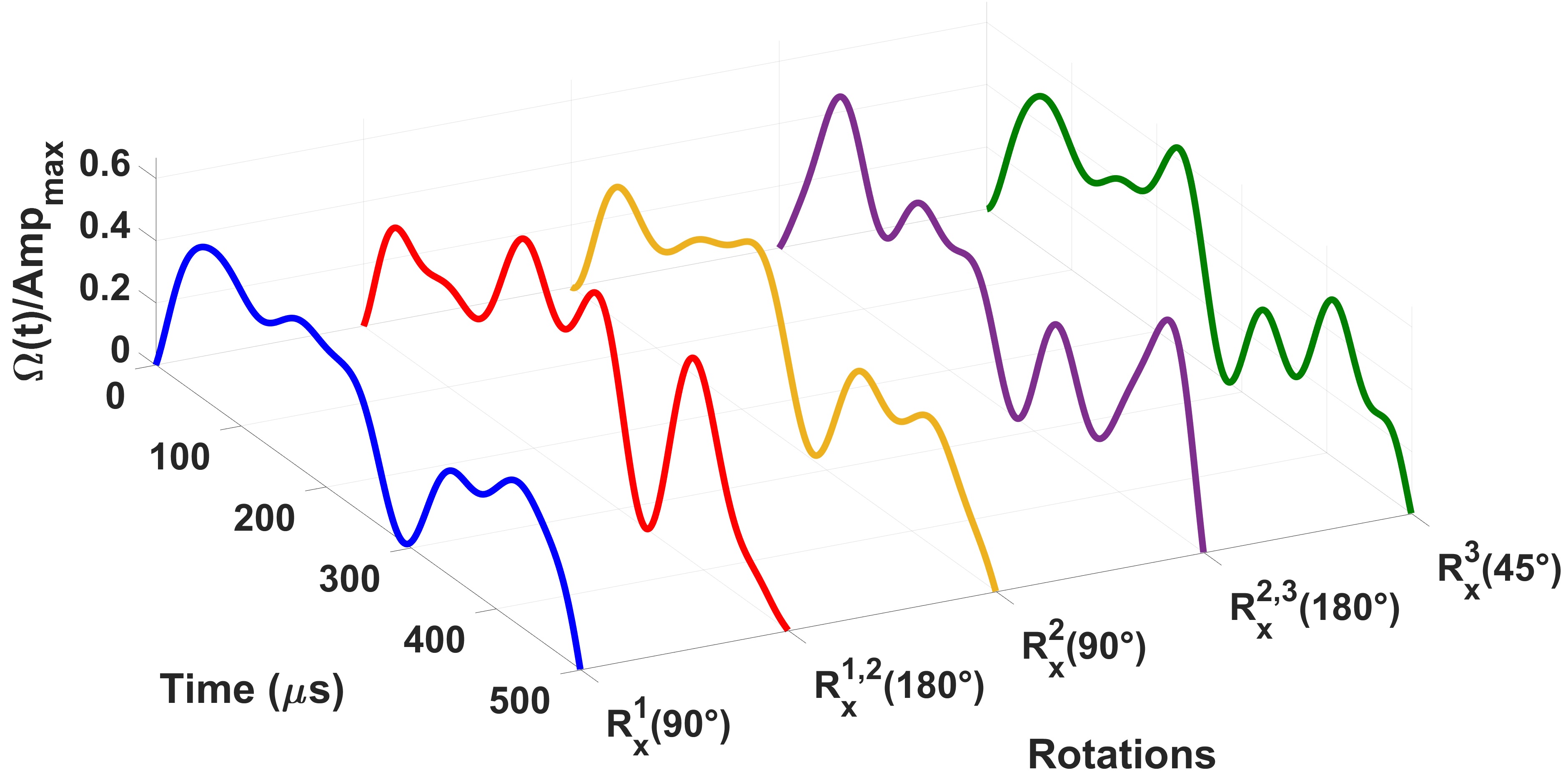} }}%
    \qquad
    \subfloat[Phase]{{\includegraphics[width=8.5cm]{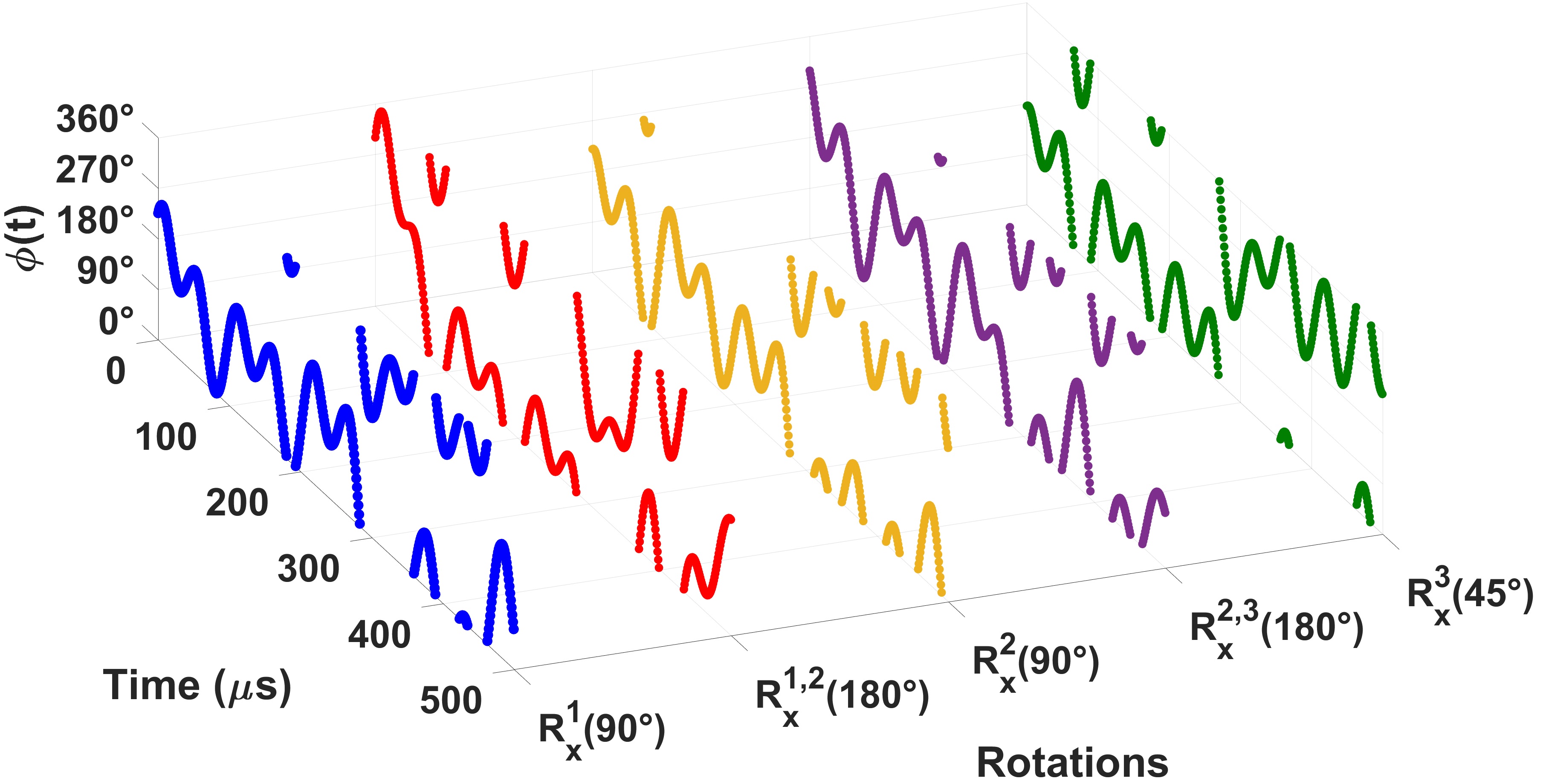} }}%
        \qquad
    \subfloat[Rf inhomogeneity]{{\includegraphics[width=8.5cm]{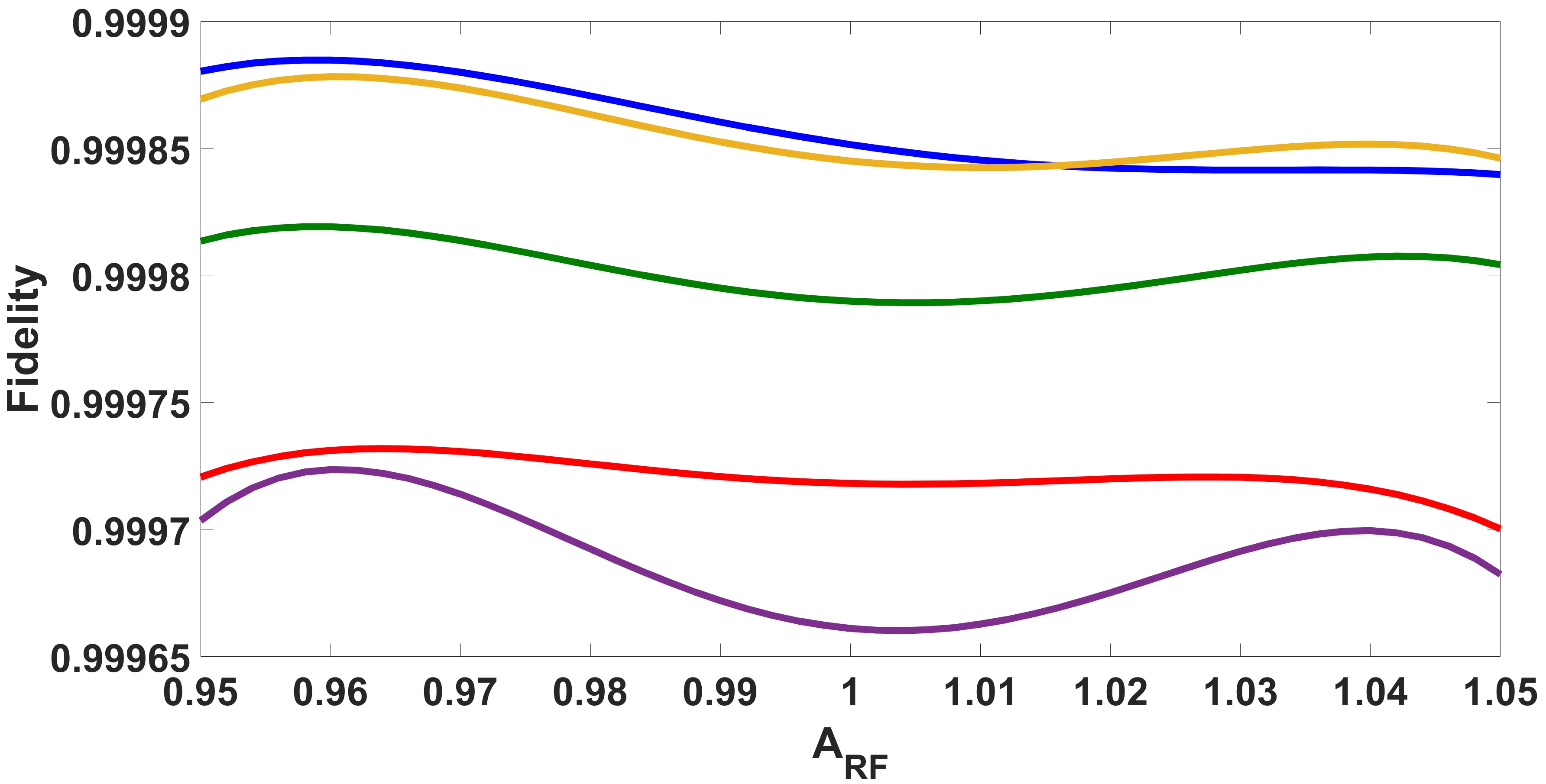} }}%
        \qquad
    \subfloat[Psuedo-pure state fidelity]{{\includegraphics[width=8.5cm]{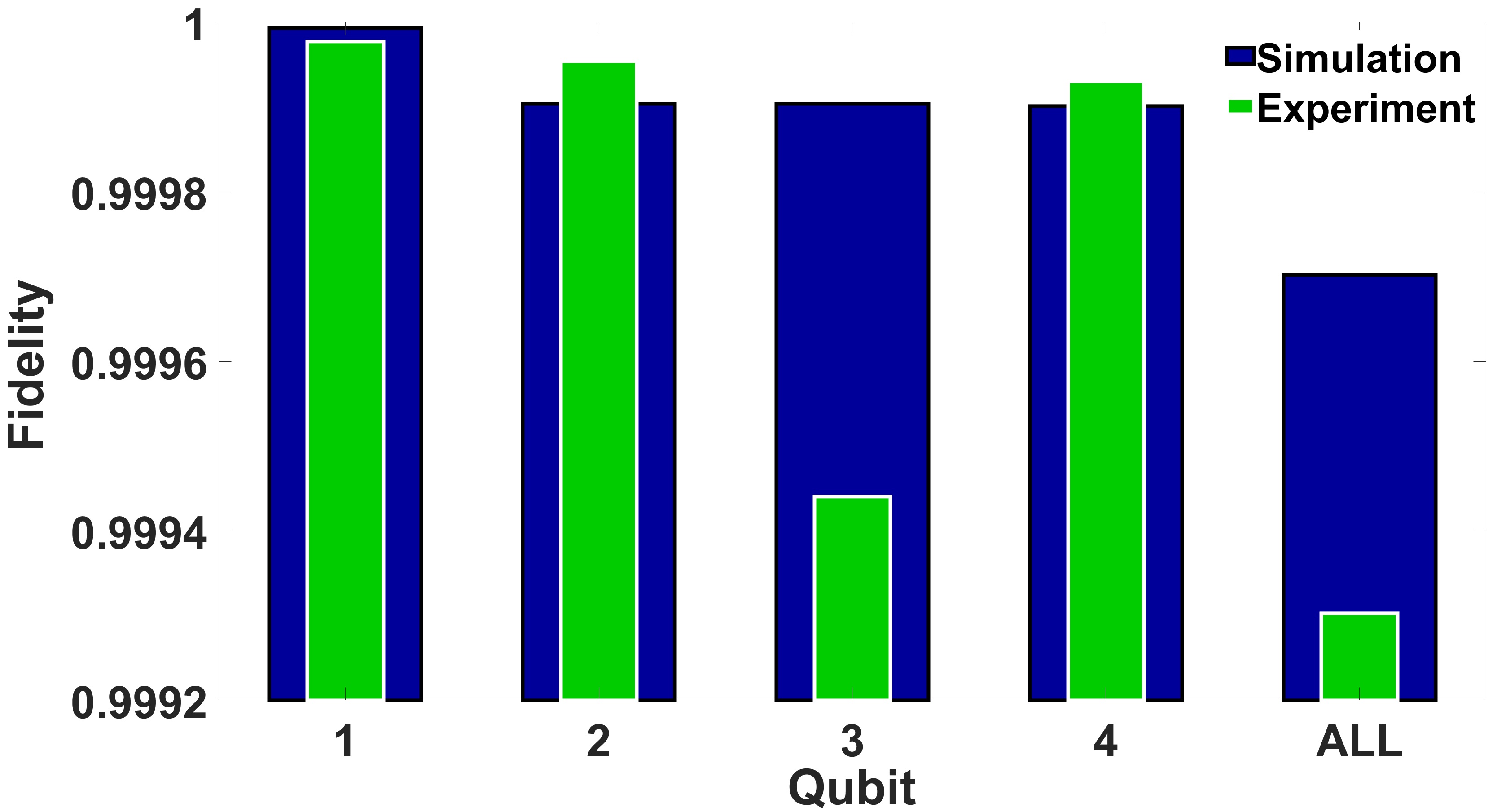} }}%
    \caption{(a-b) Modulation of the amplitude and phase of some pulses that were used to implement the rotations present in the quantum circuit shown in figure \ref{fig:pps4}. (c) Rotation fidelity when we multiply the function that describes the pulse amplitude by $\textrm{A}_{\textrm{RF}}$. (d) Fidelity obtained in the simulation and experimental implementation of the quantum circuit used to prepare the pseudo-pure state $\left | 1111  \right \rangle$. To calculate the fidelity, we determined the experimental state of each qubit using the quantum state tomography method. The last point in the graph was obtained with the calculation of the tensor product of the four experimentally determined states.}%
    \label{fig:pulsos4}%
\end{figure*}

\subsection{4 Qubits System}

\begin{figure*}[t!]%
    \includegraphics[width=17.7cm]{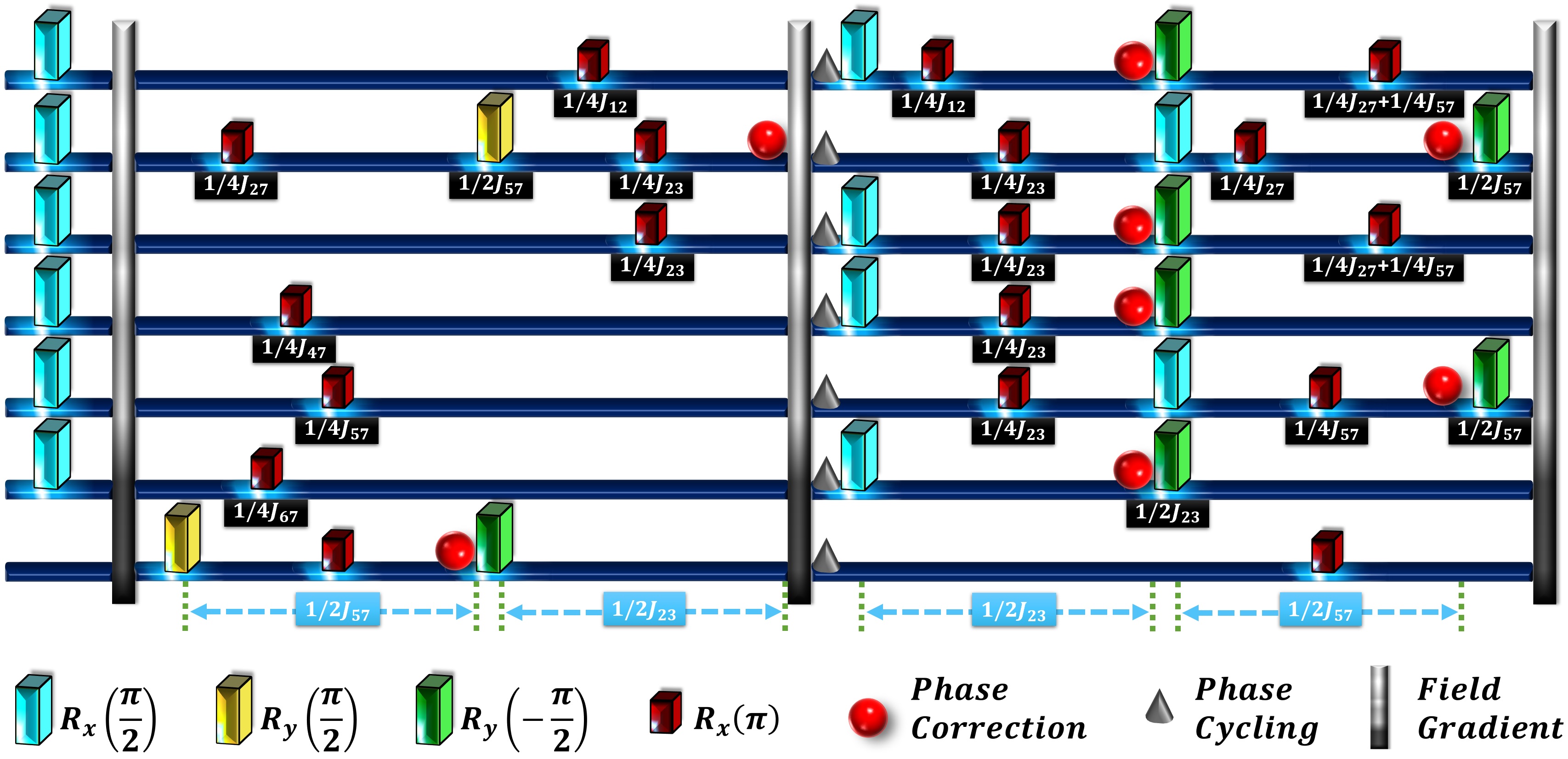}
    \centering
    \caption{ Sequence to prepare the labelled pseudo-pure state $\left | 000000  \right \rangle \left \langle 000000  \right | \sigma_{z}/2$ - In addition to rotations, we use magnetic field gradients along the $\widehat{z}$ direction and free evolution to prepare the labelled pseudo-pure state, $\left | 000000  \right \rangle \left \langle 000000  \right | \sigma_{z}/2$, starting from a thermal state. The time at which the pulse has to be applied is written below it. In the figure, $J_{kn}$ is the scalar coupling constant of the nuclear spin $k$ and $n$.}%
    \label{fig:pps7q}%
\end{figure*}

For these experiments, we used a sample of $^{13}\textrm{C}$-labelled transcrotonic acid (figure \ref{fig:molecula4q}) dissolved in acetone in order to implement the quantum circuit shown in figure \ref{fig:pps4}. In theory, with this circuit we can prepare the pseudo-pure state $\left | 1111 \right \rangle$ starting from a thermal state. In this molecule, the four $^{13}\textrm{C}$ nuclear spins, under the action of a constant magnetic field, will physically represent a four qubits system. The pulses were optimized considering that the $\textrm{H}$ nuclear spins are decoupled, this can be achieved in the experiments. The values of the resonance frequencies and the scalar coupling constants of the $^{13}\textrm{C}$ nuclear spins that were used in our algorithm are shown in figure \ref{fig:molecula4q}.

We optimized the phase and amplitude of the pulse to implement the rotations shown in the circuit of figure \ref{fig:pps4}. The free evolutions are not optimized with our algorithm, since the $\pi$ rotations and the magnetic field gradients correct most of the errors that occurs during these evolutions. Each pulse lasts $500$ $\mu$s, and 63 parameters were optimized ($s_{A} = 7$ and $s_{P} = 14$) in order to obtain $\mathcal{F}_{RF} < 0.0004$, with $\alpha_{1} = \alpha_{3} = 0.3$, $\alpha_{2} = 0.4$ and $\varepsilon = 0.05$. In our simulations we noticed that the number of parameters to be optimized can be reduced to 24 ($s_{A} = 4$ and $s_{P} = 4$), but in doing so we generally have to increase the duration of the pulse. For the 4 qubits system, we prefer to increase the number of parameters in order to reduce the duration of the pulse. When it was possible, we optimize the pulses to implement the largest number of simultaneous rotations. This reduces the total time of the experiment and, in general, improves the fidelity \cite{livronc} of the results.

In figure \ref{fig:pulsos4}(a-b), we graphically represent the amplitude and phase modulations of some of the pulses used to prepare the pseudo-pure state. We can see on figure \ref{fig:pulsos4}(c) that even when we have errors in the calibration of the pulse amplitude, the rotation will still be implemented with good fidelity. This is due to the condition we have added in our algorithm to find pulses that are robust to such errors.

After we have implemented the quantum circuit to prepare the pseudo-pure state, we determined the state of each qubit using the quantum state tomography method \cite{tomografia}. In figure \ref{fig:pulsos4}(d), we present the fidelity between the states measured experimentally and the theoretical states. We also present in this same figure the fidelity between the states obtained by simulating this circuit considering the optimized pulses and the theoretical states. When we performed the tensor product of the 4 qubits state, which were determined experimentally, and compared with the theoretical state, we found a fidelity of $0.9993$, which is exceptionally good.

\begin{figure*}[t!]%
    \includegraphics[width=17.7cm]{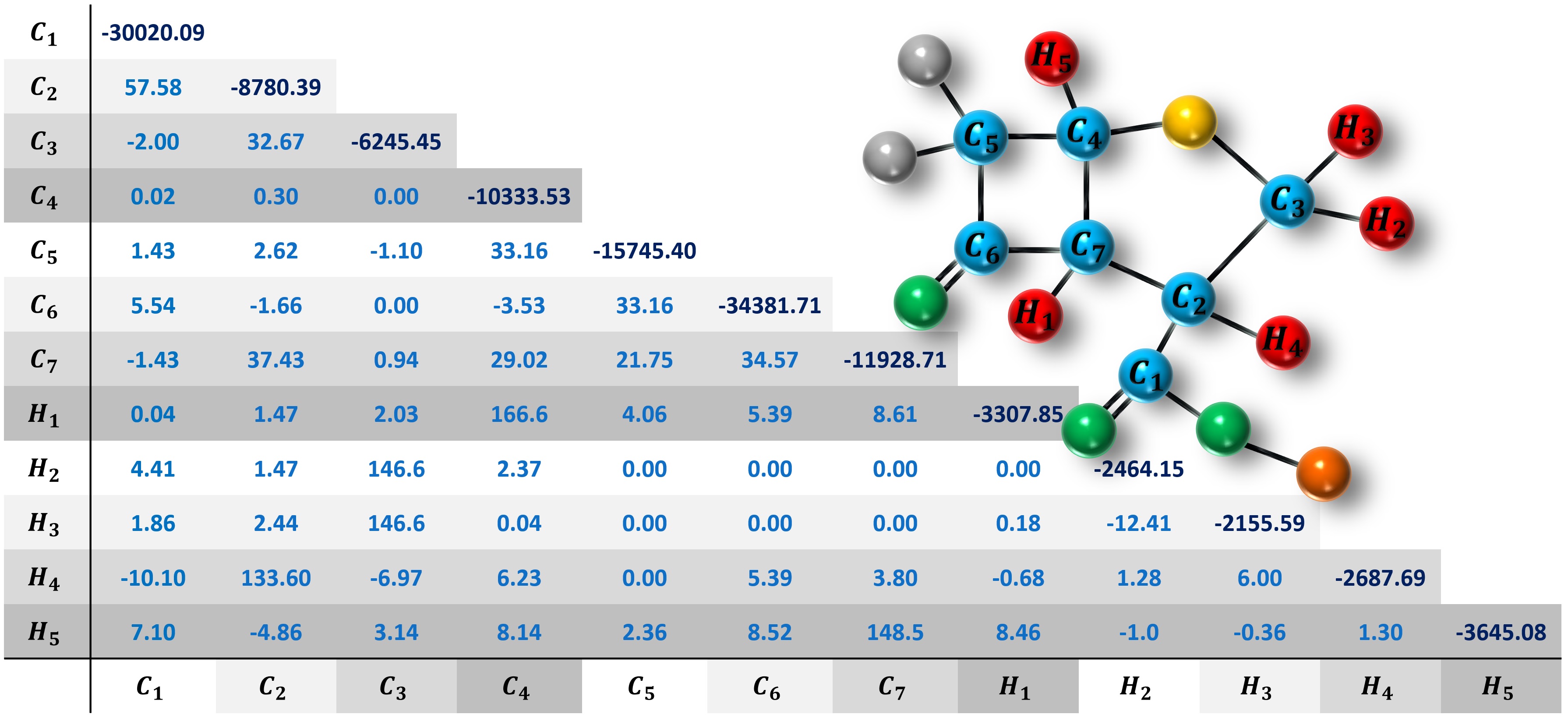}
    \centering
    \caption{ Sample information for per-$^{13}\textrm{C}$-labelled dichlorocyclobutanone molecule - The off-diagonal terms of the table are the values of the $J$ coupling constants of the $^{13}\textrm{C}$ and H nuclear spins of the per-$^{13}\textrm{C}$-labelled dichlorocyclobutanone molecule. Meanwhile, on the diagonal are written the values of the chemical shifts of each nuclear spin. The values in the table are in Hz.}%
    \label{fig:molecula12}%
\end{figure*}

\begin{figure*}[t!]%
    \centering
    \subfloat[Amplitude]{{\includegraphics[width=8.5cm]{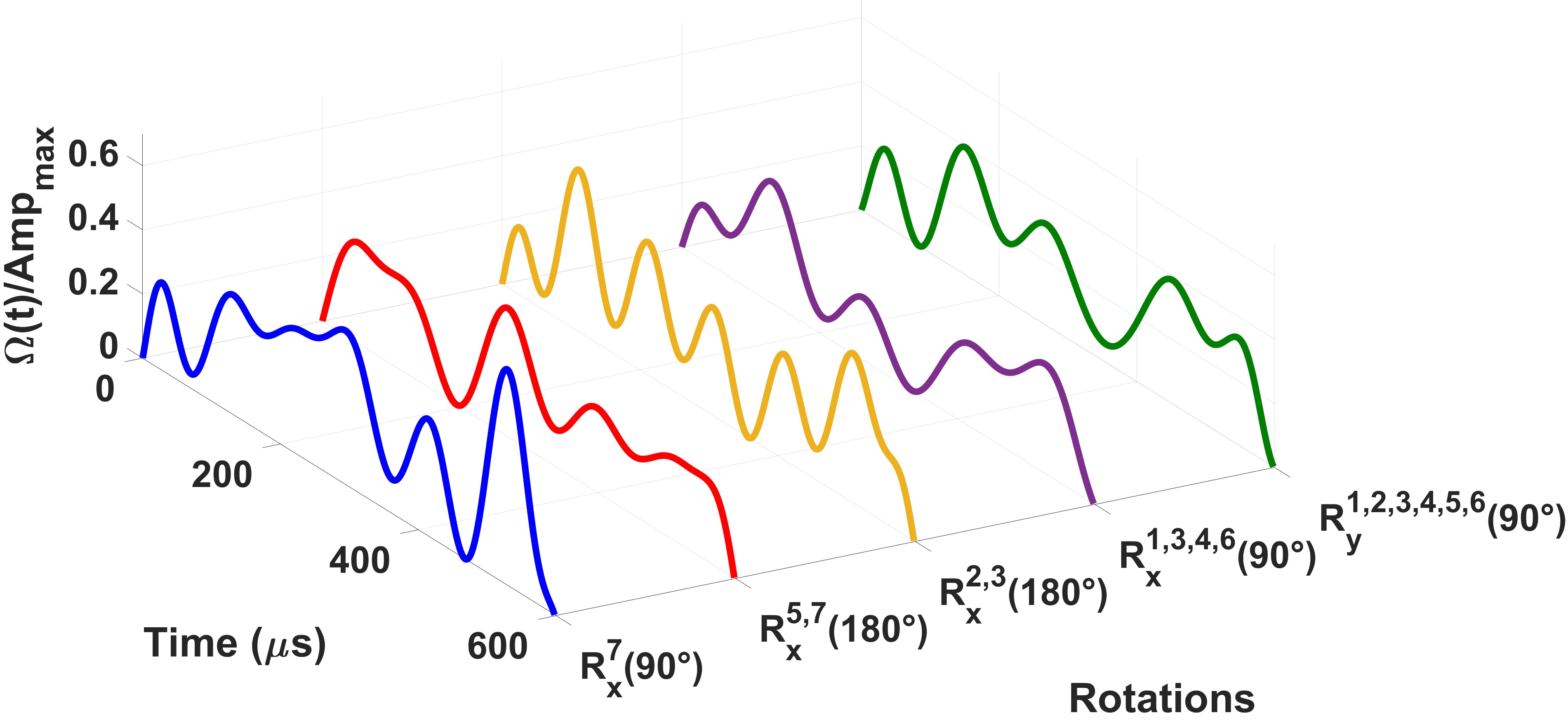} }}%
    \qquad
    \subfloat[Phase]{{\includegraphics[width=8.5cm]{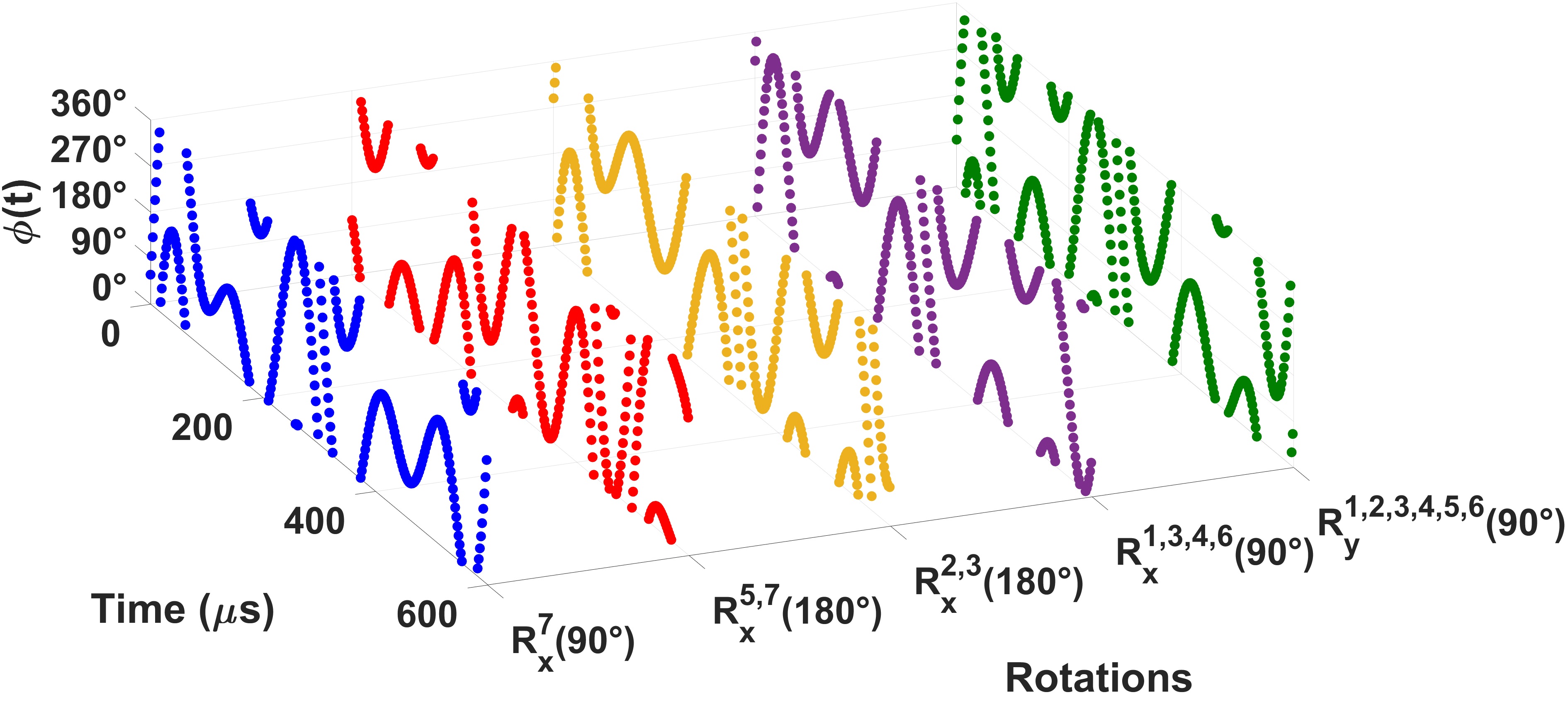} }}%
        \qquad
    \subfloat[Thermal and labelled PPS spectrum of $C_{7}$]{{\includegraphics[width=17.7cm]{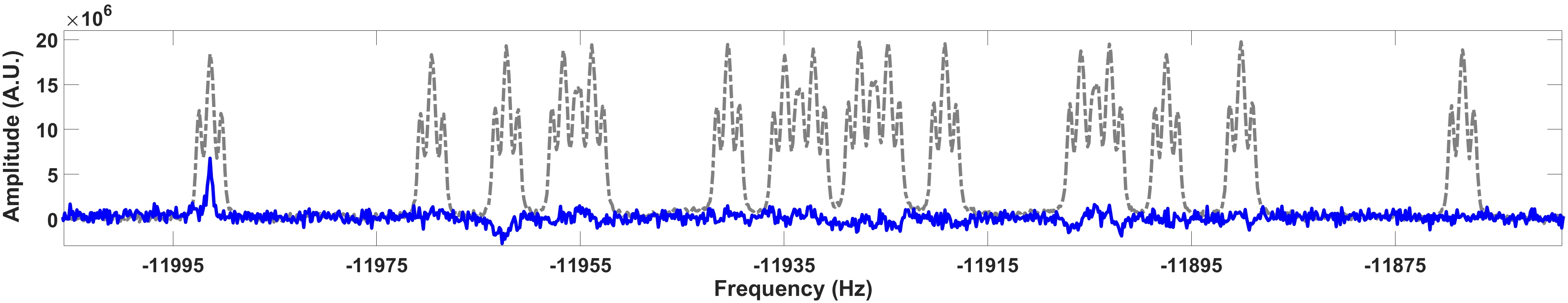} }}%
    \caption{(a-b) Modulation of the amplitude and phase of some pulses that were used to implement the rotations present in the quantum circuit shown in figure \ref{fig:pps7q}. (c) The blue line is the experimental spectrum obtained after the implementation of the sequence to prepare the labelled PPS followed by a $\pi/2$ rotation in the seventh qubit state. The gray dashed line is the experimental spectrum of the thermal state. To obtain the gray spectrum, we implemented a rotation of $\pi/2$ in the state of all nuclear spins and measured the system magnetization.}%
    \label{fig:pulsos7}%
\end{figure*}

\subsection{7 Qubits System}

We also run tests with a 7 qubit system. In this case, we use our algorithm to find the pulses that implement the rotations shown on the quantum circuit illustrated on figure \ref{fig:pps7q}. This quantum circuit is used to prepare the labelled pseudo-pure state $\left | 000000  \right \rangle \left \langle 000000  \right | \sigma_{z}/2$ \cite{lpps}, starting from a thermal state. The pulses optimization was performed considering that the nuclear spin of the hydrogen’s atoms of the per-$^{13}\textrm{C}$-labelled dichlorocyclobutanone molecule (figure \ref{fig:molecula12}) are decoupled. Thus, the nuclear spins of the $^{13}\textrm{C}$ of this molecule will physically represent a 7 qubit system. Although it is possible to divide this 7 qubit system in subgroups to accelerate the pulse optimization \cite{compila}, here we did not use this strategy, since our objective was to verify if our algorithm can provide good results as the size of the system increases.

In our simulations we used the chemical shifts and the scalar couplings shown in the figure \ref{fig:molecula12}. The amplitude and phase modulations of some of the pulses obtained with our algorithm are shown in figure \ref{fig:pulsos7}(a-b). Each pulse lasts $600$ $\mu$s and 39 parameters were optimized ($s_{A} = 5$ and $s_{P} = 8$) in order to obtain $\mathcal{F} < 0.004$. When we simulated the quantum circuit shown in figure \ref{fig:pps7q}, using the pulses optimized with our algorithm, and compared the result with the theoretical state, $\left | 000000  \right \rangle \left \langle 000000  \right | \sigma_{z}/2$, we found a fidelity greater than 0.99.

\begin{figure*}[t!]%
    \centering
    \subfloat[Amplitude]{{\includegraphics[width=8.5cm]{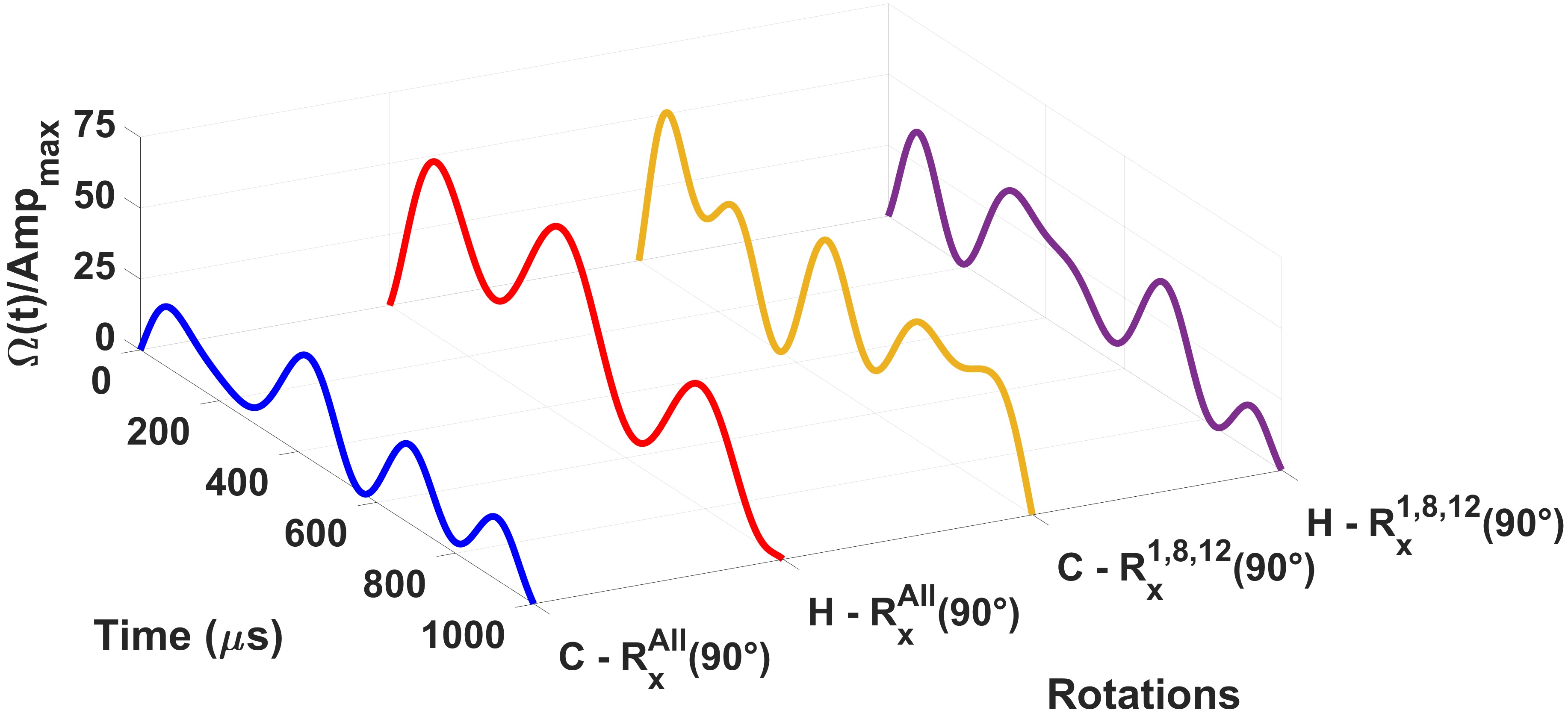} }}%
    \qquad
    \subfloat[Phase]{{\includegraphics[width=8.5cm]{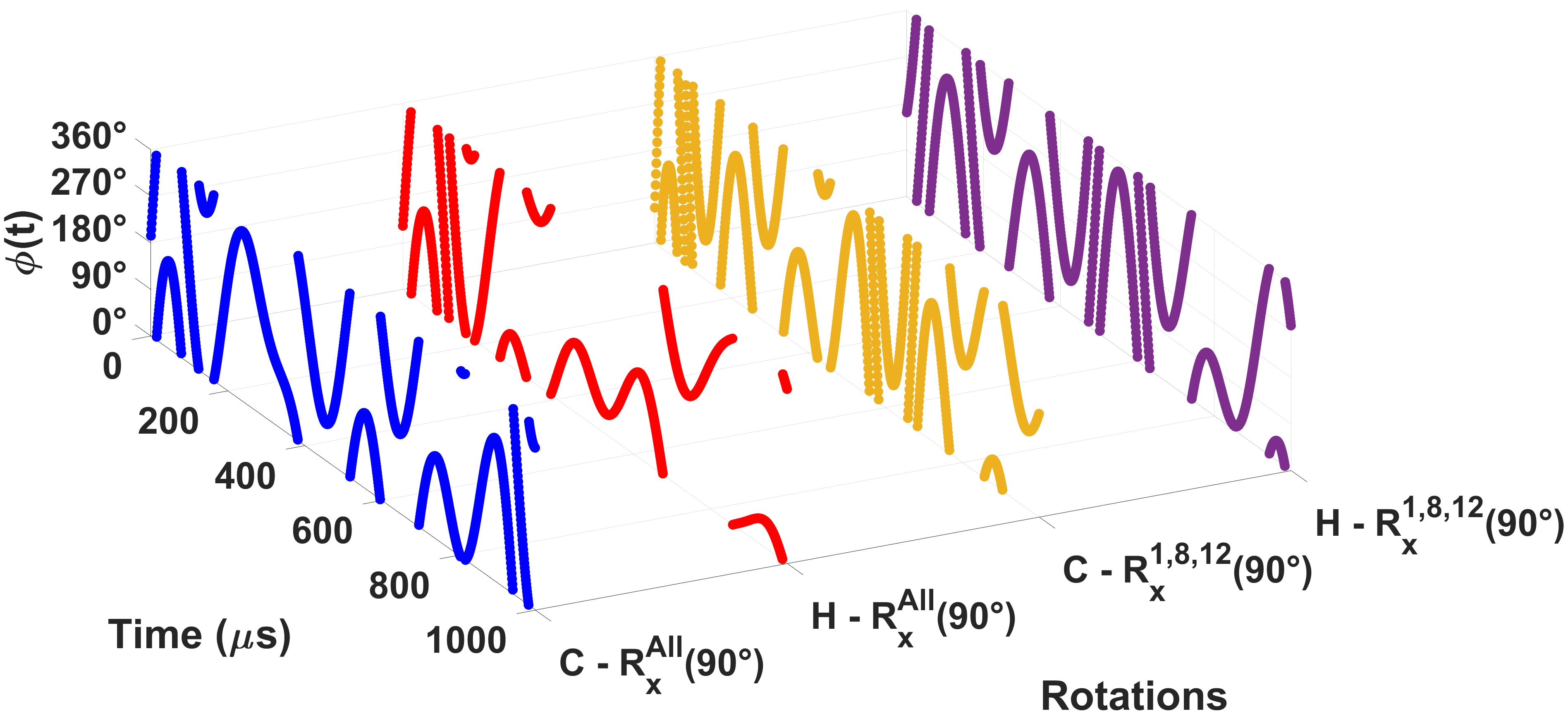} }}%
        \qquad
    \subfloat[Carbon spectrum]{{\includegraphics[width=17.7cm]{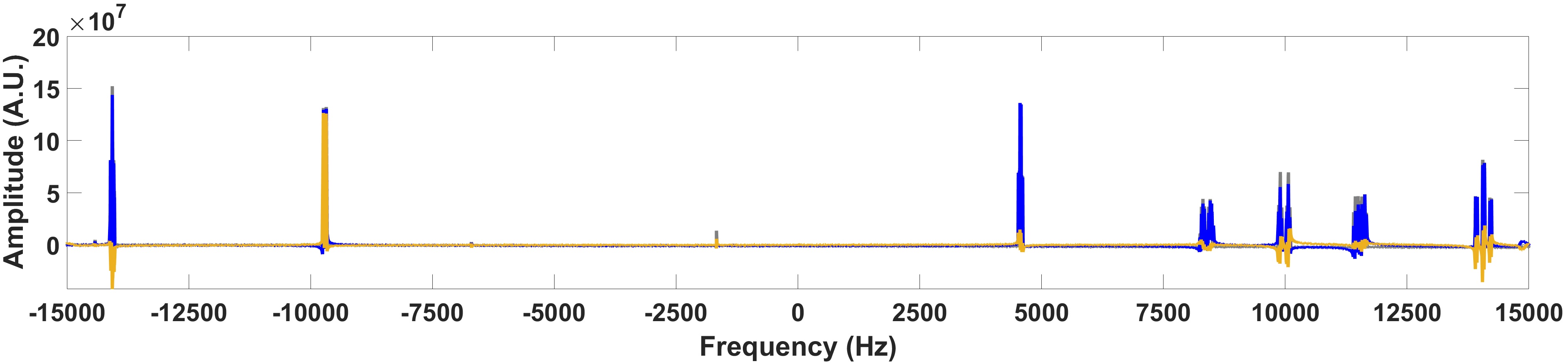} }}%
        \qquad
    \subfloat[Hydrogen spectrum]{{\includegraphics[width=17.7cm]{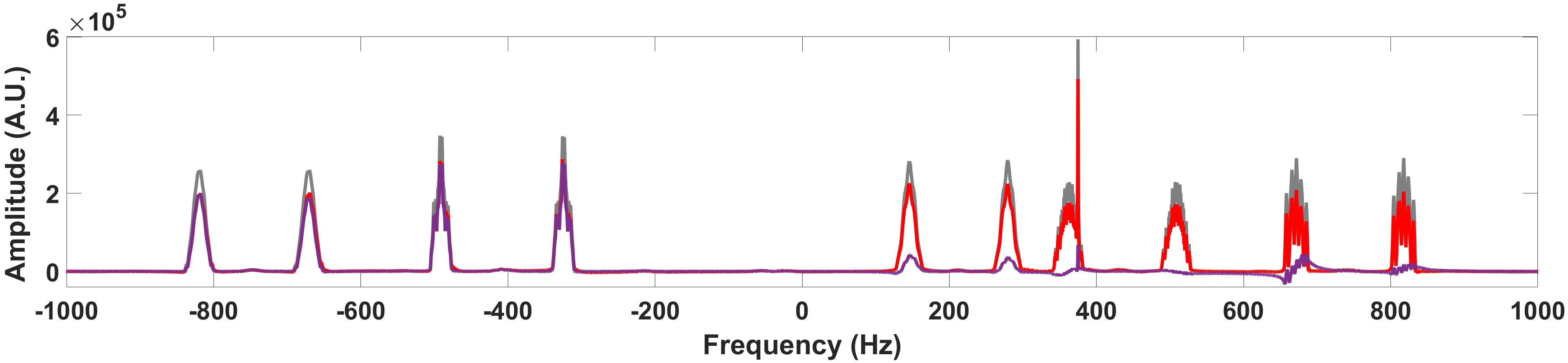} }}%
    \caption{(a-b) Modulation of the amplitude and phase of the pulses that were implemented in the 12 qubits system. In the figure, the letter C and H indicate that the pulses are applied simultaneously with a frequency close to the carbon and hydrogen resonance frequency, respectively.(c-d) The gray line is the thermal spectrum obtained by applying a fast square pulse which implements a rotation of $\pi/2$ in the state of all qubits. The other spectra were obtained by implementing the optimized pulses of (a-b) and measuring the magnetization of the system in the rotating frame.}%
    \label{fig:pulso12}%
\end{figure*}

After the pulse optimization, we implemented the quantum circuit to prepare the labelled PPS state $\left | 000000  \right \rangle \left \langle 000000  \right | \sigma_{z}/2$ and determined the state of the nuclear spin that represents the seventh qubit. Theoretically, if the system is in this labelled PPS, when we implement a rotation of $\pi/2$ in the state of the seventh qubit and measure its magnetization, we must obtain a spectrum with only one peak. In figure \ref{fig:pulsos7}(c), we present the experimental spectrum obtained after implementing the sequence to prepare the labelled PPS (followed by a $\pi/2$ rotation in the state of the seventh qubit),  along with the thermal state spectrum. As the signal coming from the sample is weak when compared to the noise, we performed 70 measurements to obtain the spectra. Even with this difficulty, we can see that it was possible to obtain a good experimental result using our algorithm to optimize the pulses. In this test we did not include the condition for the pulses to be robust to errors in their amplitude.

\subsection{12 Qubits System}

Finally, we performed some experiments with a 12 qubit system. In this case, the nuclear spins of the five hydrogen's atoms of the per-$^{13}\textrm{C}$-labelled dichlorocyclobutanone molecule (figure \ref{fig:molecula12}) were used to physically represent 5 qubits. The other 7 qubits were represented by the carbons nuclear spins. In order to optimize the pulses, we divided this system into subsystems. If we do not follow this strategy, the optimization will be slow, since it would be performed in a Hilbert space of dimension $2^{12}$. In our case, we obtained good results considering the following subsystems: $\left \{C_{1},C_{2},C_{3},H_{4}  \right \}$, $\left \{C_{2},C_{7}  \right \}$, $\left \{C_{3},H_{2},H_{3}  \right \}$, $\left \{C_{4},C_{5},C_{7},H_{1}  \right \}$ and $\left \{C_{5},C_{6},C_{7},H_{5}  \right \}$. The value of the function to be optimized, $\mathcal{F}_{sub}$, for a system composed of $n$ subsystems will be given by the following mean:
\begin{equation}\label{eq:fidesub}
 \begin{split}
\ \mathcal{F}_{sub} = \dfrac{1}{n}\sum_{k}^{n} 1 - \frac{\left | Tr\left( U_{goal_{k}}^{\dagger }  U_{k} \right)  \right |}{N_{k}},
 \end{split}
 \end{equation} 
where $U_{k}$, $U_{goal_{k}}$ e $N_{k}$ represent respectively the optimized unitary, the goal unitary and the dimension of the $k$th subsystems.

The signal coming from our sample is very weak when compared to the noise. Because of this, evaluating the results of some pulse sequences in this system can be very complicated. To work around this problem, we used only $\pi/2$ rotations in our experimental tests. Thus, in the ideal case, the spectra of the nuclear spins that are targets of these rotations must contain peaks with maximum amplitude, and the spectra of the other spins will not show peaks. In addition, to analyse the experimental results, we can use the spectrum obtained by applying a square fast pulse that implements a rotation of $\pi/2$ in the state of all nuclear spins. This fast pulse has $10$ $\mu$s of duration, and its amplitude is calibrated to obtain a spectrum with maximum amplitude.

In figure \ref{fig:pulso12}(a-b), we present the shape of the amplitude and phase of two pulses that were optimized. The first pulse implements a rotation of $\pi/2$ in the state of all nuclear spins. The second pulse implements a rotation of $\pi/2$  in the nuclear spins of the atoms of $C_{1},H_{1},H_{5}$, which represent qubits 1, 8 and 12, respectively. Each pulse lasts $1$ ms and 78 parameters were optimized ($s_{A} = 10$ and $s_{P} = 16$) in order to obtain $\mathcal{F}_{sub} < 0.007$. One of the main difficulties to optimize the pulses is the fact that the resonance frequency of the nuclear spins of the hydrogen’s atoms differ by only a few hundred Hz. Due to this fact, to individually control these spins, we had to increase the pulse duration and the number of parameters to be optimized.

After the optimization, we verified that the fidelity between the operator implemented by the optimized pulses ($U_{\mathcal{H}_{T}}$) and the ideal operator ($U_{goal}$) is greater than 0.97, by comparing the two matrices, the ideal and the calculated ones. To perform this calculation, we use Eq. (\ref{eq:uhtpro}) and (\ref{eq:uhtprok}) considering the complete system of 12 qubits to obtain $U_{\mathcal{H}_{T}}$. Due to the efficiency of our algorithm and the fact that we used subgroups, when we simulate on the same computer a pulse with discretization of $1$ $\mu$s, we note that the time required in the optimization to reach $\mathcal{F}_{sub} < 0.007$ may be 25 times less than the time required to calculate $U_{\mathcal{H}_{T}}$ of this pulse considering the complete system of 12 qubits.

In figure \ref{fig:pulso12}(c-d), we present the spectrum obtained experimentally after implementing the optimized pulses, along with the spectrum obtained by applying a fast square pulse. We can see on this figure that we have achieved good experimental results, even without including in the optimization conditions for the pulses to be robust to some types of errors. In our tests, we verified that if we double the number of parameters to be optimized, it is possible to achieve a fidelity superior to 0.995 in the simulation. However, the experimental results will not be very different from those shown in figure \ref{fig:pulso12}(c-d). In this system, even delays of a few hundred nanoseconds can affect the final result. Therefore, if we want better experimental results
in this system, we must include this and other sources of errors in our optimization. For the carbon’s nuclear spins, figure \ref{fig:pulso12}(c), we had to perform 128 measures to increase the signal-to-noise ratio. Thus, during these measurements (approximately 4 hours) the variations of temperature and magnetic field are two other sources of errors that can not be disregarded.

  \begin{figure*}[t!]%
    \centering
    \subfloat[16 qubits]{{\includegraphics[width=4cm]{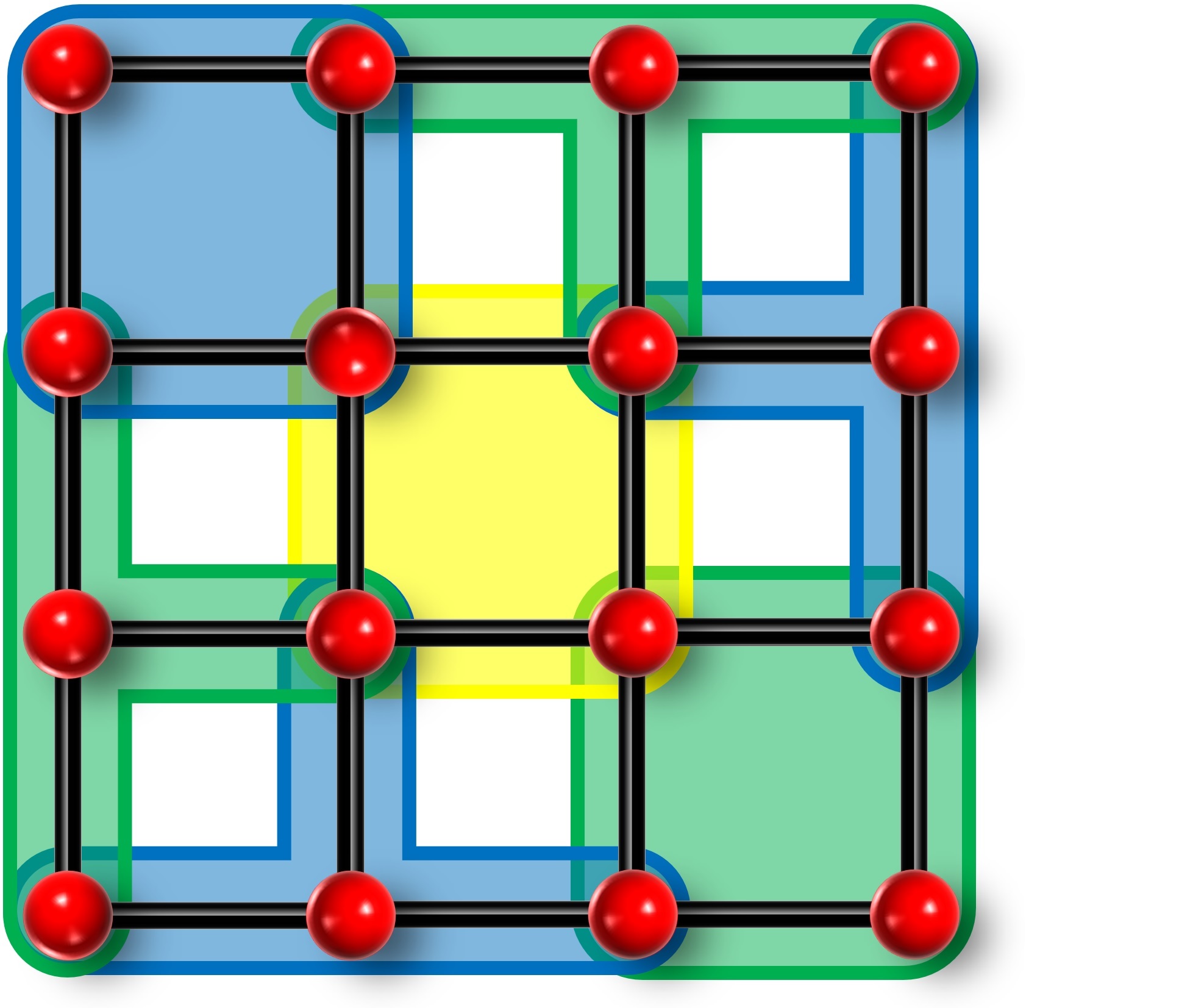} }}%
    \qquad
    \subfloat[36 qubits]{{\includegraphics[width=6cm]{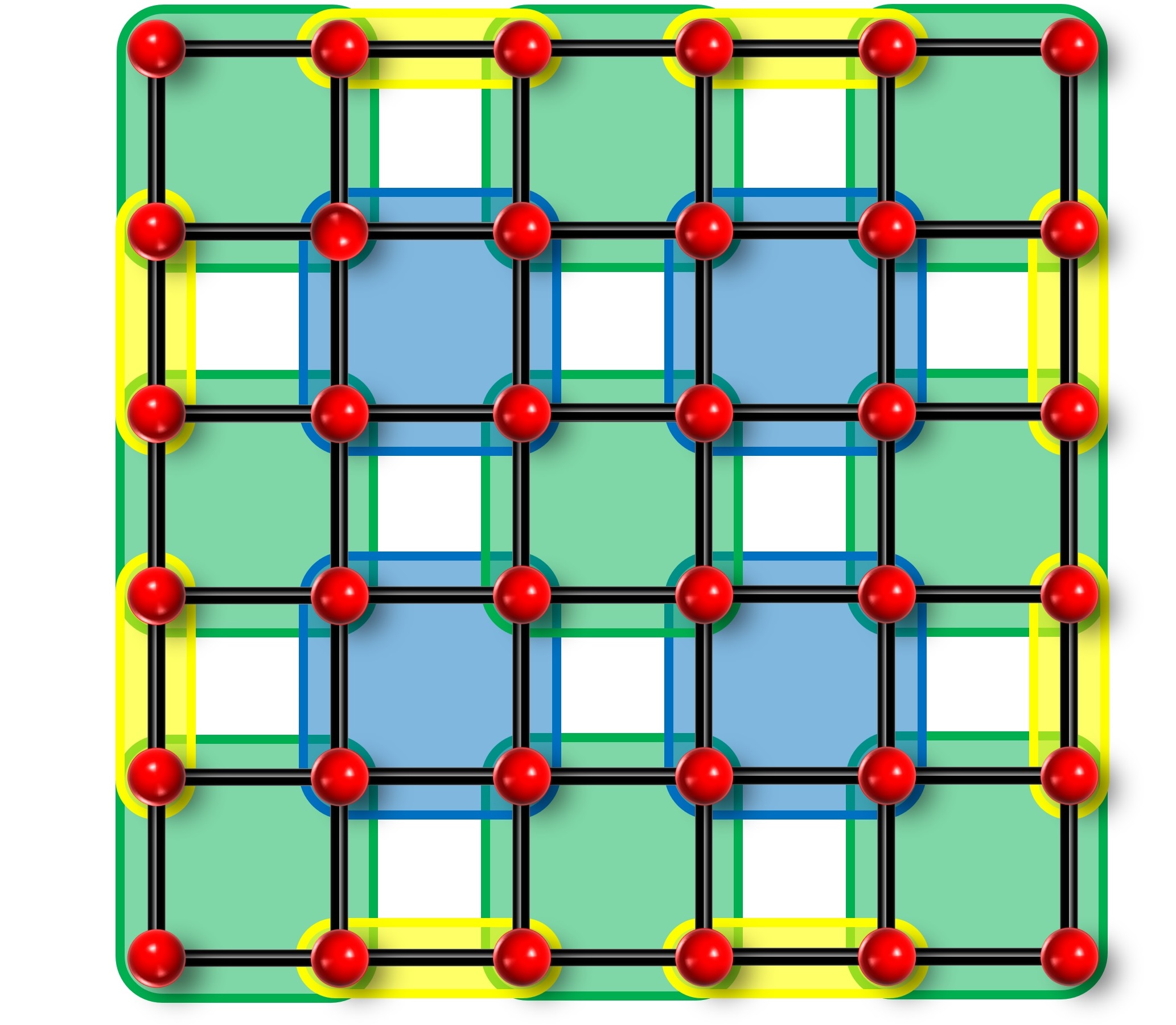} }}%
    \caption{ Square bi-dimensional lattice composed by 16 (a) and 36 qubits (b). Black bars indicate which nuclear spins pairs have non-zero scalar coupling. The subgroups used in the optimization are painted with blue, green and yellow colors. }%
    \label{fig:molecula16}%
\end{figure*}
 
\begin{figure*}[t!]%
    \centering
    \subfloat[Amplitude]{{\includegraphics[width=8.5cm]{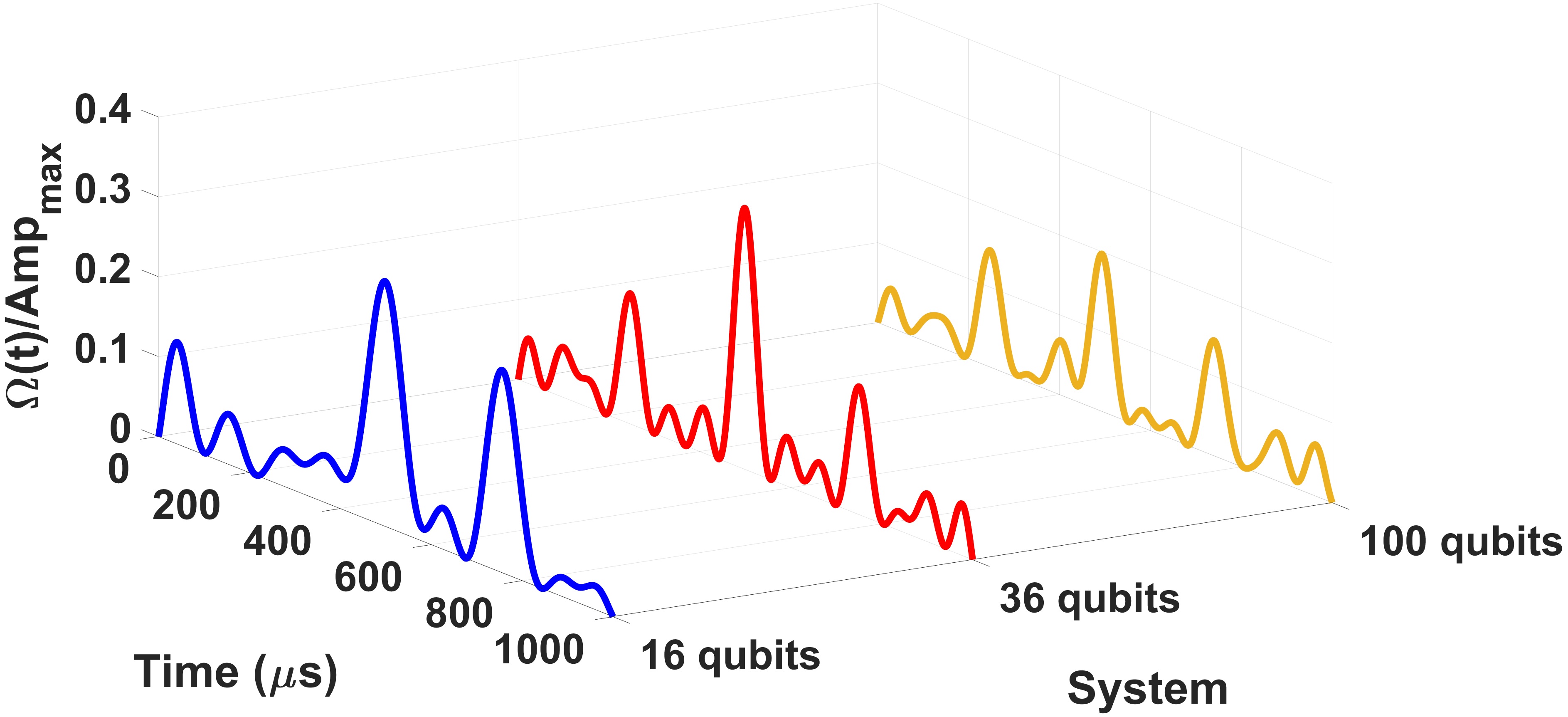} }}%
    \qquad
    \subfloat[Phase]{{\includegraphics[width=8.5cm]{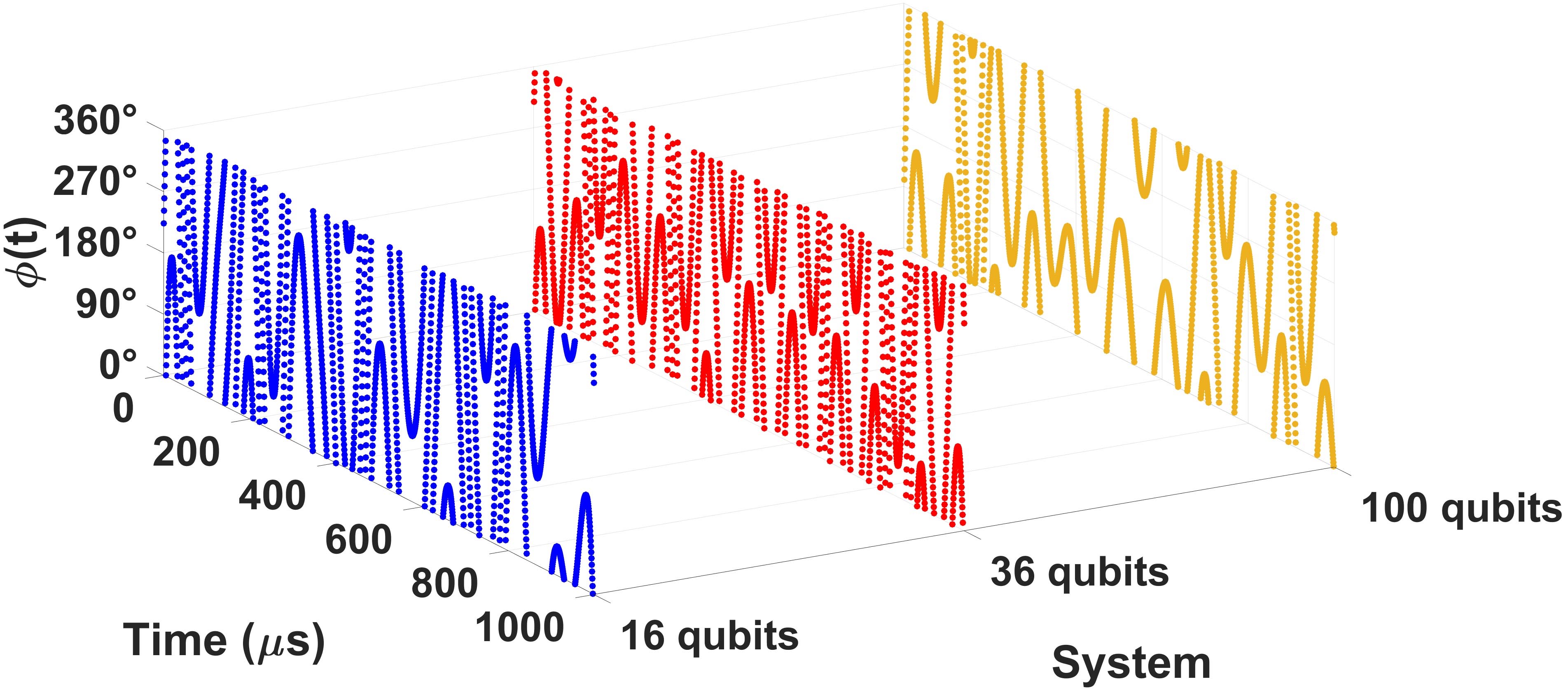} }}%
    \caption{(a-b) Modulation of the amplitude and phase of the pulses that were optimized to implement a $\pi / 2$ rotation on all odd spins of the lattice.}%
    \label{fig:pulso100}%
\end{figure*}
 
\subsection{Simulations for 16, 36, 100 and 65536 Qubits}

Our algorithm may have its performance impaired if we increase the number of parameters to be optimized, since this is based on a Nelder-Mead simplex algorithm \cite{livrooti,neldalgo,neld}. Therefore, we performed a test to verify if it is possible to optimize pulses quickly for larger systems considering the same 63 parameters ($s_{A} = 7$ and $s_{P} = 14$) used with the 4 qubits system. The systems considered are two 2D lattice with coupling between the nearest neighbours. Similar kind of systems are already an experimental reality \cite{rede2d}. It is our belief that the algorithm described in our work can be useful for controlling them.

The lattices have 16 and 36 nuclear spins, which will represent respectively systems containing 16 and 36 qubits, figure \ref{fig:molecula16}. These systems will be controlled using the NMR technique. Thus, the natural Hamiltonian of the system will be $\mathcal{H}_{0}$, Eq. (\ref{eq:h0}). The value of the scalar coupling constant ($J_{kn}$) between the nearest neighbours is equal to $50$ Hz. The angular oscillation frequency of the $k$th nuclear spin is given by: $\omega_{k} = 2\pi[b + s(k-1)]$, with $b = 700$ MHz and $s = 2$ kHz. The angular frequency of the rotating reference frame ($\omega_{R}$) is equal to the mean between the angular oscillation frequency of the first and last spin of the lattice.

In our test, we optimized the phase and amplitude of a pulse of
$1$ ms to implement a rotation of $\pi/2$ on all the odd qubits of the lattice. As in the case of 12 qubits, here we perform the optimization in subsystems. The 16 qubits system was divided into 7 groups of 4 qubits, and the 36 qubits system was divided into 13 groups of 4 qubits and 8 of 2 qubits, as shown in figure \ref{fig:molecula16}. In the optimization of the pulse in the 16 qubits system, it was possible to reach $\mathcal{F}_{sub} <0.01$ in less than an hour. Meanwhile, in the 36 qubits system, after 1 hour of optimization, we were able to obtain $\mathcal{F}_{sub} < 0.025$. The shape of the amplitude and phase of these pulses are shown in figure \ref{fig:pulso100}. To reduce the errors due to the approximation shown in Eq. (\ref{eq:uhtprokaproximado}), we decreased the value of the pulse discretization during the optimization. Initially, we started with a discretization $\delta t = 5$ $\mu$s, and we finished the optimization with $\delta t = 0.625$ $\mu$s. 

In principle, the two main difficulties to optimize pulses to control large quantum systems are the time and the amount of memory needed to perform the simulation. If we divide these systems into subgroups, we can work around those problems. For example, we did a test with a 2D square lattice with $256$x$256$, which represents a system of $65536$ qubits. The amount of memory used in the optimization process was less than 2 GB, and the time to calculate all the quantities needed to obtain the value of $\mathcal{F}_{sub}$, Eq. (\ref{eq:fidesub}), was approximately 1 minute (considering a pulse of $1$ ms and $\delta t = 10$ $\mu$s). In our simulations, we used a computer with an Intel Core i7-8700 processor and 16 GB of RAM. The subgroups were divided following the same pattern used for the 36 qubit system (figure \ref{fig:molecula16}). While it is possible to work with such large systems using our algorithm, when we increase the size of the lattice without changing the number of optimized parameters, the maximum amplitude or duration of the pulse, we may have difficulty obtaining $\mathcal{F}_{sub} < 0.01$ in the optimization of the pulses. This happens because the distance between the resonance frequency of the first and last spin of the lattice also increases. It is worth remembering that the error of the approximation presented in Eq. (\ref{eq:uhtprokaproximado}) will also increase when this distance increases. One way to solve this problem is to consider that the lattice is formed by more than one kind of atom and that we can use multiple rotating frames \cite{livro}. In this case, the resonance frequency of the atoms of a species may be several hundred MHz different from the frequency of the atoms of the other species. It is worth mentioning that when the difference between the pulse frequency and the resonance frequency of the nuclear spin increases, it will be more difficult for this pulse to change the state of the spin

In our simulation we consider a spectrometer with 5 channels (5 rotating frames), as used in \cite{sonda}. As was done previously, we used 63 parameters ($s_{A} = 7$ and $s_{P} = 14$) to optimize the amplitude and phase of a pulse of $1$ ms to implement a rotation of $\pi /2$ on all the odd qubits of the lattice. In this case, the pulses applied in the 5 channels will have the same shape for the amplitude and phase but their frequencies of oscillations will be different. We performed the simulation assuming the lattice is composed of 100 qubits. The lattice is composed by nuclear spins of atoms $^{1}\textrm{H}$, $^{19}\textrm{F}$, $^{13}\textrm{C}$, $^{31}\textrm{P}$ and $^{15}\textrm{N}$. The lattice is configured to have the first 20 qubits represented by nuclear spins of $^{1}\textrm{H}$, the next 20 by nuclear spins of $^{19}\textrm{F}$ and so on. In each group of 20 spins, the resonance frequency of the $k$th nuclear spin of the group will be given by: $\omega_{k}^{n} = 2\pi[b_{n} + s(k-1)]$, with $s = 2$ kHz and $b_{n}$ represents the characteristic resonance frequency of the nuclear spin of the $n$th species of atom of the lattice. If we consider an NMR equipment with a magnetic field magnitude of $16.44$ T, we have $b_{H} = 700$ MHz, $b_{F} = 658$ MHz, $b_{C} = 176$ MHz, $b_{P} = 283$ MHz and $b_{N} = -71$ MHz. In order to perform the optimization, we divided the system following the same pattern used for the 36 qubits system (figure \ref{fig:molecula16}). After approximately 3 hours of optimization, it was possible to obtain $\mathcal{F}_{sub} < 0.012$, with a $\delta t = 0.625$ $\mu$s, which is an excellent result. The shape of the amplitude and phase of the optimized pulse is illustrated in figure \ref{fig:pulso100}. It is worth remembering that this pulse was optimized considering the same restrictions, used in the optimization of the pulses for a system of 7 and 12 qbits, for it to be well implemented experimentally. It is possible to conclude from the results that even with a small number of parameters, our algorithm can efficiently optimize pulses for these big systems considered here. Since our algorithm can be fast and does not require a large amount of memory, we believe that it can contribute significantly to the control of large quantum systems that could be used as quantum computers, not only in NMR but also for other technologies.

\section{Conclusions}

In summary we have developed an algorithm for optimizing radio-frequency pulses, generally used in NMR systems in order to implement quantum gates with high fidelity. The pulses can be optimized to be robust to calibration errors. Besides, with our algorithm we can obtain pulses that have smooth modulations, since these pulses are described by a set of smooth functions. This is a good advantage over some others methods, since most NMR spectrometers do not deal well with fast variations of the pulse parameters. These functions can be chosen experimentally to ensure that the optimized pulses are implemented with good precision. Additionally, in the method we have developed a small number of parameters are used and consequentially the whole optimization process is performed faster than in other methods. We have shown the success of our algorithm using real NMR experiments, where systems composed of 4 to 12 qubits where controlled.  Finally, we have proved that, even in a system with 100 qubits, the pulses used to implement rotations can be described by a small number of parameters, and our algorithm can be efficient enough to optimize a modulated pulse, in a short period of time. Thanks to the effectiveness of our optimization algorithm, we were able to obtain good experimental results without using any other error correction technique or external calibration devices. Ascribed to this effectiveness and efficiency of our algorithm, we believe that it can be used to control large quantum systems in other experimental techniques, others than NMR. A future challenge would be to employ this algorithm to control systems containing thousands of qubits.

\begin{acknowledgments}
We thank Hemant Katiyar and Janet Venne for valuable discussions that helped to develop the ideas presented in this paper. We thank Shayan-Shawn Majidy for valuable comments on the manuscript. We acknowledge financial support from Ministery of Innovation,
Science and Economic Development (Canada), the Government of Ontario, CIFAR, Mike and Ophelia Lazaridis, CNPq and FAPERJ.
\end{acknowledgments}

\end{document}